\documentstyle[11pt,paspconf,psfig]{article}
\newcommand\rosat{{\it ROSAT}}
\newcommand\hst{{\it HST}}
\newcommand\asca{{\it ASCA}}

\newcommand\ha{H$\alpha$}
\newcommand\hii{H{\small II}}
\def\gtrsim{\lower 2pt \hbox{$\, \buildrel {\scriptstyle >}\over
{\scriptstyle \sim}\,$}}
\def\lesssim{\lower 2pt \hbox{$\, \buildrel {\scriptstyle <}\over
{\scriptstyle \sim}\,$}}

\begin{document}

\title{Hot Gas and Physical Structure of 30 Dor}

\author{Q. Daniel Wang}
\affil{Dept. of Physics \& Astronomy, Northwestern University, 2145 Sheridan Road, Evanston,~IL 60208-3112}

\begin{abstract}
Recent \rosat\ and \asca\ X-ray observations as well as \hst\ images 
have provided new insights into the structure and evolution of the 30 Dor nebula.
I review hot gas properties of the nebula and discuss related physical 
processes. The structure of the nebula can be understood as
outflows of hot and \hii\ gases from the parent giant molecular cloud of R136. 
The dynamic mixing between the two gas phases is likely a key mass loading 
process of the hot gas, providing a natural explanation of both temperature 
and density structure of the nebula. 
\end{abstract}


\section{Introduction}

	Hot gas in the 30 Dor nebula was first detected by the {\sl Einstein}
Observatory (Wang \& Helfand 1991; Chu \& Mac Low 1990). The nebula apparently 
consists of blisters of hot gas surrounded by \ha-emitting filaments. Stellar 
winds and supernova explosions are presumably responsible for the
heating of the gas. Recently, the nebula has been observed 
with instruments  on board the \rosat\ and \asca\ X-ray Observatories. 
X-ray images from some of these observations have been presented 
(Chu 1993; Wang 1995a; Wang \& Gotthelf 1998). But detailed analysis of the 
data is still ongoing. In this writing-up of my talk, I 
review some preliminary results on hot gas properties and discuss 
implications on various relevant physical processes.

\section{Hot Gas Properties}

	Fig. 1 shows a state-of-art high resolution X-ray image of the
30 Dor nebula (Wang 1998). In its core region, two point-like sources with 
0.1-2.4~keV luminosities of $\sim 10^{36} {\rm~ergs~s^{-1}}$ stand out and 
are most likely Wolf-Rayet + black hole binary systems (Wang 1995a). 
The presence of such systems indicates that supernovae have occurred in the 
recent past of 30 Dor. The X-ray contribution from the central star cluster 
R136 is also detected (Fig. 2). The image shows 
no evidence for other point-like X-ray sources which may be related to 30 Dor.
The detection limit is $\sim 10^{35} {\rm~ergs~s^{-1}}$. There is also no 
correlation between the X-ray intensity and the stellar distribution. The 
X-ray emission is thus predominantly diffuse in origin.
A three color composite image of the nebula in \ha, UV, and X-ray can be found
at http://www.astro.nwu.edu/astro/wqd/images/30d.htm. Wang \& Gotthelf (1998)
have presented an overlay of X-ray contours on an \ha\ image and \asca\ SIS 
data in two energy bands. In addition, Chu (1993) has shown a \rosat\ PSPC 
image of the nebula. 

\medskip
\centerline{
\hfil\hfil
\psfig{figure=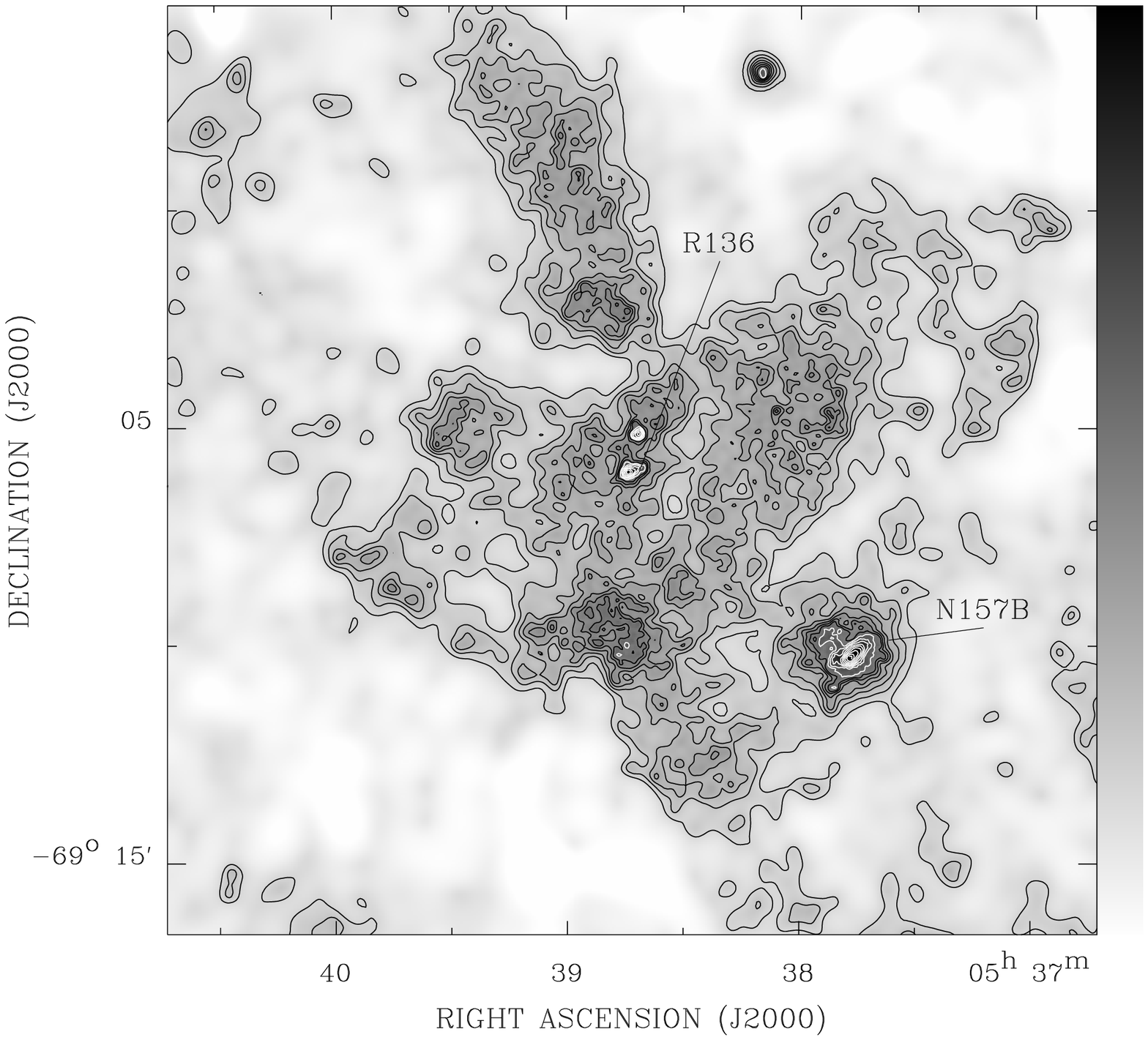,height=2.5truein,angle=90.0,bbllx=58bp,bblly=128bp,bburx=550bp,bbury=680bp,clip=}
\psfig{figure=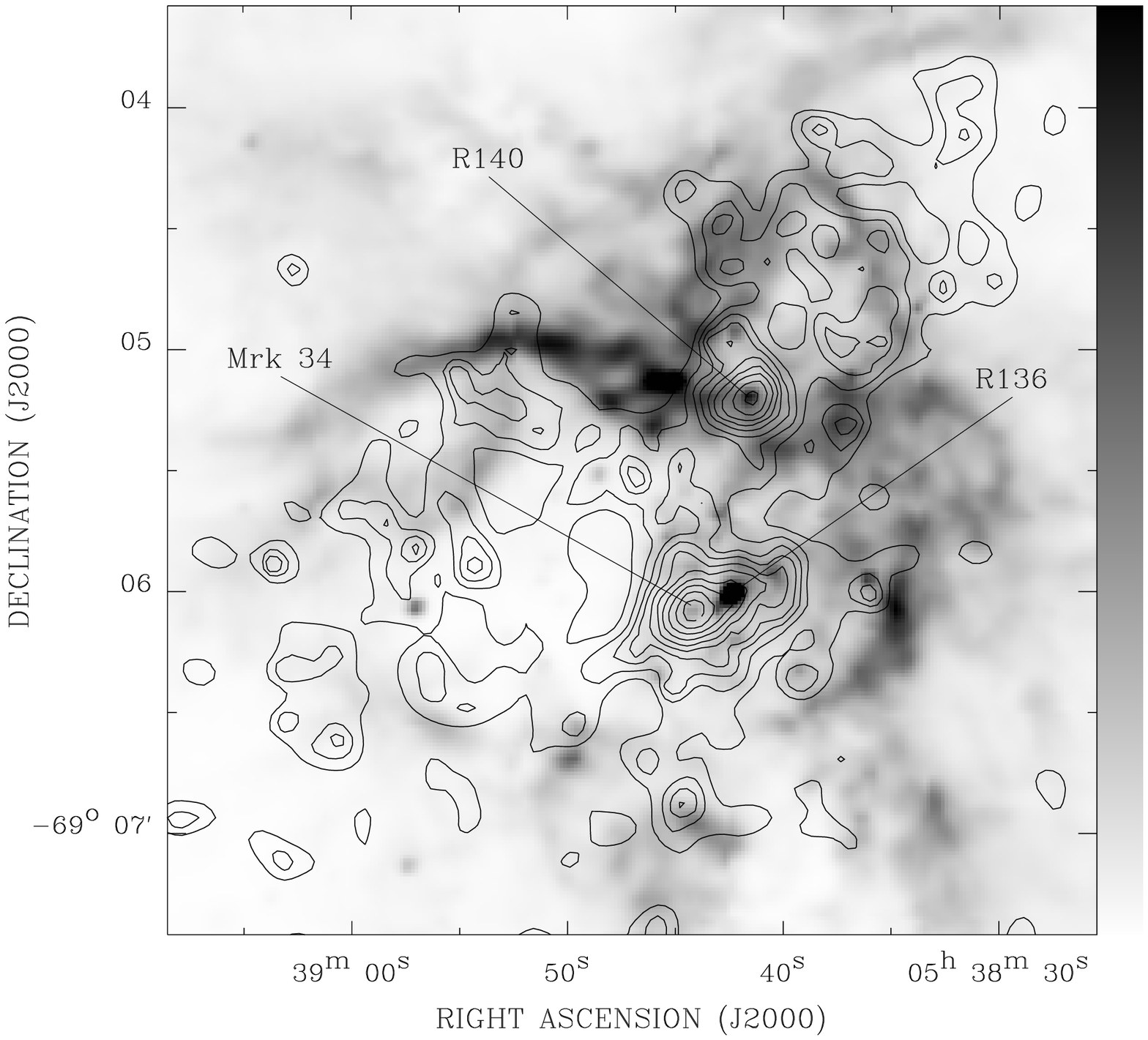,height=2.5truein,angle=90.0,bbllx=58bp,bblly=128bp,bburx=550bp,bbury=680bp,clip=}
}

\begin{figure}[h]
\caption{(left panel) X-ray image of 30 Dor. The image is a co-add of data 
from \rosat\ HRI exposures of total 130~ks. The contours are at 
0.6, 1.2, 1.9, 3.0, 4.5, 6.4, 9.1, 15, 24, 39, 63, 100, 160, 256, 407, and 
$647 \times 10^{-3} {\rm~counts~s^{-1}~arcmin^{-2}}$, above a local 
background of $\sim 1 \times 10^{-3}$.} 
\caption{(right panel) The central region of 30 Dor in X-ray and \ha. 
The RHRI contours start at $ 3 \times 10^{-3} {\rm~counts~s^{-1}~arcmin^{-2}}$.
The rest is the same as Fig. 1.
} 
\end{figure} 

	These observations show that the 30 Dor
nebula is an interacting complex of massive stars, dense molecular clouds, 
photoionized \hii\ filaments, and diffuse hot gas. Young massive stars are concentrated in R136. But a few smaller clusters
within the central region ($r \lesssim 4^\prime$) may also contribute
significantly to the energetics of the nebula. Dense molecular clouds 
typically appear in regions that are relatively dim in soft X-ray, apparently
a result of displacement and/or X-ray absorption. The central region, in 
particular, shows hot gas bubbles surrounded by ionization fronts at boundaries of
dense molecular clouds (Fig. 2).

	An archival \asca\ observation (Fig. 3) provides X-ray
spectroscopic information about hot gas in the 30 Dor nebula (Wang 
1998). The presence of individual emission lines (e.g., Ne, Mg, and Si) 
confirms that X-ray emission in the 0.5-2~keV band is predominantly thermal 
in origin. The flat spectrum at higher energies, however, suggests 
the presence of a nonthermal component. A power law
of energy slope equal to one describes this component well, 
although the exact spectral
shape is uncertain because of the limited energy coverage and statistics
of the data. Using the Raymond \& Smith thermal plasma to characterize the 
thermal portion of the spectrum, I find that at least two temperature 
components
are needed for a reasonable good fit. These two components characterize a
range of hot gas temperature 0.2-1 $\times 10^7$~K in the nebula. The high
temperature component dominates in the energy band $\gtrsim 1.5$~keV, and 
arises primarily in the core region around R136 ($\lesssim 1^\prime$;
Wang \& Gotthelf 1998). The total inferred thermal energy and mass of 
hot gas are $\sim 10^{52.5}$~ergs and $10^{4.6} M_\odot$, contained primarily 
in the low temperature component.

\begin{figure}
\centerline{
\psfig{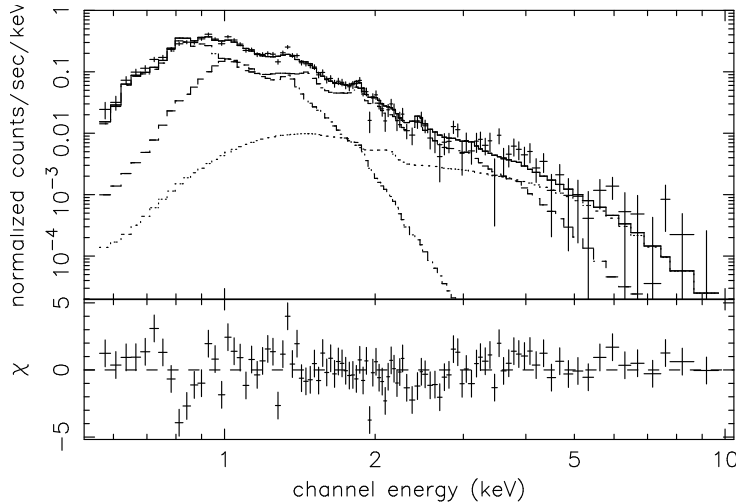}}
\caption{Average \asca\ SIS spectrum of the X-ray emission from 30 Dor.
A three-component model fit is shown as the solid histogram. 
The low and high temperature thermal plasma components are plotted as 
dot-dashed and dashed histograms, while the power law as dotted histogram.} 
\end{figure} 

\section{Physical Processes}

	 The thermal energy in the hot gas is balanced by the expected 
mechanical energy release from stellar winds and supernovae in the nebula.
The stellar winds from R136 alone has an energy release rate of 
$L_w \sim 3 \times 10^{39} {\rm~ergs~s^{-1}}$ (e.g., Chu \& Kennicutt 1992).
Over the lifetime of R136, the integrated energy release is 
$\sim 10^{53}$~ergs, a large fraction of which should have been converted 
into the thermal energy of hot gas. There could also be tens of supernova 
explosions, although the exact number is difficult to estimate.

	The thermal structure of the nebula is directly related to the cooling 
of hot gas. The radiative cooling rate of the hot gas is about 
$\sim 10^{39} {\rm~ergs~s^{-1}}$, a reasonable fraction of the 
estimated energy input rate. The shocked stellar wind has a 
temperature $\sim 10^8$~K and is not effective in X-ray emission. 
As hot gas moves from the core to the halo of the nebula,
adiabatic cooling may be important. The average pressure $p/k$ is about
$ \sim 2 \times 10^7 {\rm~K~cm^{-3}}$ in the core and is a factor
of $\sim 7$ lower in the halo. However, the corresponding adiabatic 
temperature drop of a factor $\sim 2$ is  not adequate to explain the
temperature difference in the two regions. The adiabatic
expansion should also rarefy the gas by a factor of $\sim 3$. 
But the measured mean hot gas densities are comparable in the two regions. 
Therefore, mass loading needs to be considered.

	The most effective mass loading process is likely the dynamic mixing 
of hot gas with \hii\ gas. While the mass efficiency of 
massive star formation is typically only 5-10\% (e.g.,  McKee 1989), much of 
the parent giant molecular cloud (GMC) of the OB association NGC 2070 is being 
eroded via photon-evaporation. As the 30 Dor nebula has entered the so-called 
champagne phase (Wang 1995b), the photon-evaporated \hii\ mass can be 
estimated as (Whitworth 1979; York et al. 1989) 
$\sim 0.12 M_\odot {\rm~yr^{-1}}$$ (S_{i}/10^{52} 
{\rm~Ly~photons~s^{-1}})^{4/7} $$
(t_\ast/10^6 {\rm~yr})^{2/7}$$ (n_o/10^3 {\rm~cm^{-3}})^{-1/7},$
where $S_i$, $t_\ast$, and  $n_o$ are the ionizing flux of the OB association, 
its effective age with the present flux, and 
 the mean density of the GMC. Only if $\sim 10\%$ of this photon-evaporated
\hii\ gas is loaded to the shocked wind materials in
the core of the nebula, can the temperature of hot gas there be explained.
Over the lifetime of the nebula, the total evaporated mass is 
$\sim 10^5 M_\odot$. In order to account for the total mass of hot gas, 
the mass loading over the nebula must be substantial.

	The dynamic mixing is evident, right within 
the central cavity around R136. The cavity is clearly open to the east. The 
size of the cavity, as manifested by ionization fronts,
is about half arcminute radius, or $R \sim 7$~pc.  Multi-filter
\hst\ WFPC2 images (Scowen et al. 1998 and this volume) clearly demonstrate 
the presence of evaporative flows from the ionizing fronts into the interior of
the cavity. Such flows result from the strong pressure gradient 
(a factor of $\sim 10$)
between the ionization fronts and the interior of the cavity. 
Both the interaction between 
stellar winds and evaporative flows and the interface between the outflows 
of \hii\ and hot gases are highly unstable. The dynamic mixing is 
bound to occur. Supernova blastwaves can further sporadically heat large amounts
of \hii\ gas. The high pressure in the core region tends to drive the heated 
gas out to the halo, leading to the blister-like morphology of 
the nebula.

	Clearly, the study of 30 Dor has strong implications for 
the understanding of giant \hii\ regions in general. The early mass loading,
in particular, may determine hot gas properties of supergiant 
bubbles and galactic chimneys, which are all evolved from giant \hii\ regions 
(Wang 1995b).


%
%

\begin{question}{Dr.\ G. Snonneborn}
What is known about the kinematics of the hot gas?
\end{question}
\begin{answer}{Dr.\ Q. D. Wang}
Unfortunately, the available X-ray instruments don't yet have the spectral
capability to yield any information about the kinematics. In the near future,
useful information may be obtained from FUV spectroscopic measurements of 
elements such as O{\small VI}, which are sensitive to gas at temperatures of
a few times $10^5$~K.
\end{answer}

\begin{question}{Dr.\ J. Bregman}
	Hunter told us that the R136 cluster has an age of 1-3 Myr and that
the massive stars have not evolved into SN. However, your picture requires
that tens of SN have gone off. Can you reconcile these apparently contradictory
results?
\end{question}
\begin{answer}{Dr.\ Q. D. Wang}
R136 is the main central cluster of 30 Dor. But the central region contains
a quite few other star clusters, which appear to be older. The presence of
the X-ray binary systems does indicate SN in the region. There are
other pieces of evidence for SN, as You Hua has argued. Energetically, 
my picture does not really require any SN, within the 
uncertainties of the present measurements.
\end{answer}

\begin{question}{Dr.\ H. Zinnecker}
I would like to make a comment. From your number of the hot, X-ray-emitting
gas ($n_e = 0.2 {\rm~cm^{-3}}, T = 5 \times 10^6$~K), we can calculate the hot
gas pressure ($\sim 10^6~k_B$). This is 10 times higher than the internal 
pressure of typical dense cold proto-stellar globules ($n = 10^4 {\rm~cm^{-3}}, 
T = 10$~K). So the X-ray-emitting gas can indeed trigger secondary star formation
in the 30 Dor region (cf. N. Walborn review of the generation of young stars
revealed by NICMOS/HST observations).
\end{question}

\end{document}